\documentclass[journal]{IEEEtran}
      
\PassOptionsToPackage{bookmarks=false}{hyperref}
  
\usepackage{cite}  
\usepackage{amsmath,amssymb,amsfonts}
\usepackage{algorithmic}
\usepackage{graphicx}
\usepackage{subfig}      
\usepackage{multirow} 
\usepackage{textcomp} 
\usepackage{hhline} 
 
\begin{document}  
  
\title{Memoryless Techniques and Wireless Technologies for Indoor Localization with the Internet of Things
}

\author{Sebastian~Sadowski,~\IEEEmembership{Student Member,~IEEE},
                Petros~Spachos,~\IEEEmembership{Senior Member,~IEEE}, and Konstantinos~N.~Plataniotis,~\IEEEmembership{Fellow,~IEEE} 
\thanks{This work was supported in part by the Natural Sciences and Engineering Research Council (NSERC) of Canada.
\par S. Sadowski and P. Spachos are with the School of Engineering, University of Guelph, Guelph, ON, N1G 2W1, Canada.
(e-mail: ssadowsk@uoguelph.ca; petros@uoguelph.ca).
\par K. Plataniotis is with the Department of Electrical and Computer Engineering, University of Toronto, Toronto, ON, M5S3G4, Canada. (e-mail: kostas@ece.utoronto.ca)}}
  
\maketitle
\thispagestyle{empty}
\pagestyle{empty}
 
\begin{abstract} 
In recent years, the Internet of Things (IoT) has grown to include the tracking of devices through the use of Indoor Positioning Systems (IPS) and Location Based Services (LBS). When designing an IPS, a popular approach involves using wireless networks to calculate the approximate location of the target from devices with predetermined positions. In many smart building applications, LBS are necessary for efficient workspaces to be developed. In this paper, we examine two memoryless positioning techniques, K-Nearest Neighbor (KNN), and Naive Bayes, and compare them with simple trilateration, in terms of accuracy, precision, and complexity. We present a comprehensive analysis between the techniques through the use of three popular IoT wireless technologies: Zigbee, Bluetooth Low Energy (BLE), and WiFi (2.4~GHz band), along with three experimental scenarios to verify results across multiple environments. According to experimental results, KNN is the most accurate localization technique as well as the most precise. The RSSI dataset of all the experiments is available online.


\end{abstract}

\section{Introduction}
\IEEEPARstart{R}{ecent} advancements in the Internet of Things (IoT) have lead to the emergence of new applications, one of which being positioning, commonly known as localization. Localization in its simplest terms is the process of making something local to an area, which can be achieved through the use of Indoor Positioning Systems (IPS) and Location Based Services (LBS). IPS are used in calculating the target's location, while LBS use the target's location information to control features of the environment.
 
Most localization systems often require real-time information from devices with known positions, referred to as anchors, to accurately calculate the location of an unknown device~\cite{lashkari}. In many outdoor localization systems, Global Positioning System (GPS) is often used in computing the positional information of a desired device. However, when indoors, multipath effects and complex environments can cause large errors in the position calculations~\cite{multipath1, multipath2}. Hence, in order to determine a position indoors, other methods need to be used. Currently, many types of solutions exist, but no standards are in place for an IPS. 

Due to the multitude of challenges faced in localization, designing a single system that is capable of tracking devices in all types of environments with  high accuracy is not feasible. Since there exist many types of environments with a variety of layouts, sizes, and obstacles, a system designed to function in one location might not function at all in another. Therefore, when designing an indoor localization system there is no trivial solution. In order for optimal results to be determined, advanced knowledge of the area where localization is occurring is required. 

In this paper,  we compare the performance of two memoryless positioning techniques and three  wireless technologies in three different environments. For the analysis, the techniques chosen were  K-Nearest Neighbor (KNN), and Naive Bayes as well as  trilateration. Trilateration was selected due to its simpleness of implementing using Received Signal Strength Indicator (RSSI) readings and its popularity in localization systems. On the other hand, KNN and Naive Bayes were chosen due to their popularity in indoor localization systems utilizing fingerprinting. Three wireless technologies were selected in order to verify results with different system designs. The wireless technologies selected were Zigbee, Bluetooth Low Energy (BLE), and WiFi (2.4 GHz band). These technologies were selected based on their presence in smart city scenarios, popularity in IoT applications, and their ease in measuring the RSSI of received signals. The experiments took place in three environments with different sizes and different interference levels, low, average, and high. For the room with low interference, there were  no other transmitting devices in the area. For the room with high interference, there were four other devices configured to use the same wireless channels with the experimental devices and for the  room with average interference, a laboratory with a number of computers  and users in the area was used. The RSSI dataset that was built from the experiments is available online~\cite{RSSIdata}.

The main contributions of this work are as follows:
\begin{itemize}
\item Systems utilizing commercially available hardware were developed comparing multiple localization techniques compromising of multiple wireless technologies.
\item Extensive experimentation was performed, building a dataset of RSSI data~\cite{RSSIdata}.  The experiments took place under three scenarios to verify results across rooms with different dimensions and interference.
\item A detailed analysis is presented which determined that fingerprinting with KNN processing is a highly accurate technique for localization.
\end{itemize}

The rest of this paper is organized as follows: the related work is reviewed in Section~\ref{related}, and the localization techniques in Section~\ref{techniques}. The wireless technologies selected are discussed in Section~\ref{metho}, with the experimental scenarios being discussed in Section~\ref{scenarios}. The experimental setup and process are described in Section~\ref{exp}, followed by the results and discussion in Section~\ref{res}. Section~\ref{con} concludes this work.

\section{Related Work} \label{related}
 When designing a localization system, the technique being used to calculate the estimated location is an important part of the system. While most systems utilize trilateration due to its ease and scalability~\cite{wifi_trilateration, csi_trilateration}, others shift toward fingerprinting as it can provide a much higher accuracy at the cost of a larger system set up time~\cite{wang, yiu, song, knnFormula, knnformula_2}. Comparisons are performed between several designs to determine which would be optimal for the intended application~\cite{knnformula_2, techniques, comp_fingerprinting, mobile_comparison}. Since applications can greatly differ from one another including the intended deployment sites, the designed systems should be able to function based on the specifications provided. It is important that a proper selection is made for the different parameters of a system. In~\cite{techniques}, a comparison was performed between three fingerprinting techniques: Bayesian, KNN, and Neural Networks. Results demonstrated that KNN produces the highest accuracy, but required the longest running time to compute a position. In~\cite{comp_fingerprinting}, a comparison between three fingerprinting algorithms, Bayes, Euclidean, and Isoline, was presented. Results concluded that Isoline produced the highest accuracy. Technology has greatly advanced in recent years and further experimentation with recent approaches is needed. There is a need for an up-to-date comparison between techniques and wireless technologies to determine which would be optimal for an indoor localization system using fingerprinting.

In~\cite{fingerprinting_algorithms}, existing WiFi access points were used in a comparison between the RSSI fingerprinting algorithms: Euclidean distance, Pompeiu-Hanusdorff Distance, and Kullback-Leibler Distance, using both weighted and non-weighted KNN processing. The proposed methodology created a large set of results through the use of multiple testbeds along with experimental scenarios. Results demonstrated that algorithms with lower complexity produced results with greater accuracy than those with higher complexity. Results also showed that between small and large environments KNN provided a minimal increase in the accuracy.
  
In~\cite{mobile_comparison}, a comparison was performed between two techniques, multi-trilateration, and Nearest-Neighbor, while running on a mobile device using WiFi access points. Results proved that Nearest-Neighbor had a much higher accuracy when compared to multi-trilateration. However, experiments demonstrated that using fingerprinting required a larger amount of computational resources, resulting in calculations having a lower latency compared to multi-trilateration. Having a lower latency would not be acceptable in a real-time localization system as having a delay to compute the results would not be optimal in the tacking of objects. Results also proved that as the number of reference points increased, the accuracy decreased. By only selecting reference points with the lowest measured RSSI values out of all the available points, the overall accuracy of the system could be increased.

In addition to the localization algorithm, selecting a wireless technology to use is also an important factor that needs to be taken into account in a localization system~\cite{spachos4}. Based on popularity, WiFi is the most commonly used wireless technology among localization systems~\cite{rssi2, models, wifi_trilateration2, knnformula_2, comp_fingerprinting, mobile_comparison}. With the recent development of BLE, a large number of systems are now focusing on using BLE beacons~\cite{blespachos} for indoor localization purposes~\cite{rezazadeh, example_ble, ble_fingerprinting, spachos3}. 
Due to its low power consumption and popularity in IoT applications, Zigbee, while not as common as WiFi or Bluetooth, has been seeing an increased amount of use for localization purposes~\cite{zigbee1, zigbee2, zigbee3}. Each technology has each own advantages and disadvantages when it comes to be used to localization. Currently, a lack of research has occurred in comparing the different technologies to determine which would be ideal and the most accurate for a localization system.

In this paper, we expand on the works discussed above and compare two memoryless techniques:  KNN, and Naive Bayes, along with simple, trilateration, utilizing three experimental scenarios to demonstrate the functionally of the systems under multiple locations. To verify results, three wireless technologies, Zigbee, BLE, and WiFi (2.4 GHz band) were used to determine how the accuracy is affected and which would be the preferred technique to use for an indoor localization system.

\section{Localization Techniques} \label{techniques}

In the design of an indoor localization system, most types of systems are either model-based or survey-based~\cite{models}. In model-based systems, locations are determined mathematically through calculations utilizing the distances or angles between transmitters and receivers. In survey-based systems, environments are analyzed in detail before the system is deployed by building a database containing the areas of interest. Among commonly used techniques for indoor localization are lateration~\cite{trilateration} and fingerprinting~\cite{fingerprinting}, the former being model-based and the latter being a survey-based system.

\subsection{Trilateration}
Trilateration is a model-based technique where the receiver's location is determined mathematically through the use of distances. To calculate using trilateration three transmitting devices are necessary to find a 2D position and four are required to find a 3D position. In addition to the proper number of transmitting devices, the distances between the transmitters and the receiver are required. In order to calculate the distance between the devices, a popular approach includes the RSSI of a signal. The RSSI of received signals is readily available to devices, therefore it is a low-cost and effective method to use for indoor localization.  A detailed explanation of the trilateration process followed can be seen in~\cite{mine}.

RSSI is a measurement of the power present in a received signal. Most often RSSI is used for determining the quality of a signal on a device from an access point. Shorter distances between devices produce signals with  high quality, hence,  larger RSSI values. As the distance between devices increases the signal becomes weaker and the RSSI drops. 

Unfortunately, the RSSI measurements are often unreliable. Signals can interfere with each other, obstacles can cause Non-Line-of-Sight~(NLoS) conditions, and reflections of objects can cause the signal's strength to fluctuate causing a poor localization accuracy. All of these factors are important to consider and can make performing localization with trilateration difficult. Since fingerprinting requires the building of a database, a lot of the factors that affect wireless signals are not as present, hence fingerprinting is referred to as a more accurate technique.

\subsection{Fingerprinting}
In order to apply a fingerprinting technique in the deployment of a localization system two phases are required. In the first phase, sampling occurs by measuring RSSI values at points of interest in the environment and storing the locations of those points at a database. In the second phase, when an unknown location needs to be mapped, the RSSI values at the unknown location are sent to the database where it is compared with the stored values through an algorithm. It then returns a match with the most likely position that the device is located. Algorithms that have been applied in localization systems utilizing fingerprinting include KNN using Euclidean distance~\cite{knnJournal}, Expectation-maximization~\cite{expectation-maximizationJournal}, Gaussian process~\cite{yiu}, Neural Networks~\cite{neuralnetworkJournal}, and Bayesian Estimation~\cite{fingerprinting_algorithms2}.  Fingerprinting techniques are known to be highly accurate in localization systems, however, mapping an environment can be time consuming and labor intensive~\cite{models}. If any changes were to occur in the environment, the database would need to be recreated since RSSI values are heavily influenced by obstacles, reflections, and multipath effects.

\begin{figure}[t!]
\centering
\includegraphics[width=0.6\linewidth]{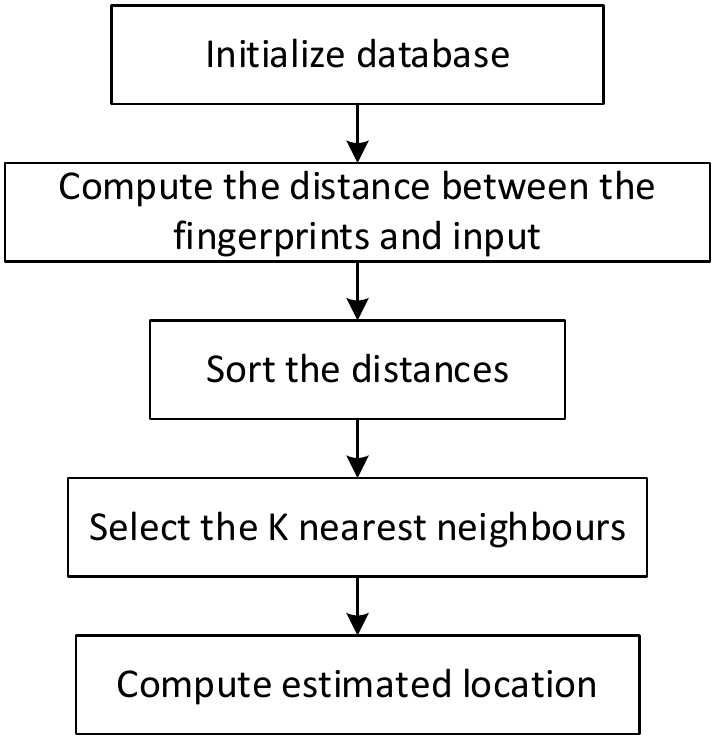}
\caption{KNN processing algorithm flowchart.}
\label{fig:knnflowchart} 
\end{figure}

\subsubsection{K-Nearest Neighbor}
KNN can be implemented for comparison between the RSSI values when fingerprinting is used. A flowchart detailing the KNN process is shown in Fig.~\ref{fig:knnflowchart}.   In a simple KNN algorithm, measured RSSI values from access points at an unknown location are compared with the RSSI values stored in the database using the Euclidean Distance~\cite{knnFormula}, as follows:

\begin{equation}
\label{eq:knn}
D_i = \sqrt[]{\sum_{j=1}^{n}(RSSI_{ij} - RSSI_j)^2}, \quad i=1,2,3,\dots,M
\end{equation}
where \(D_i\) is the distance between the measured RSSI value (\textit{\(RSSI_j\)}) at access point \textit{j} of a test location and the recorded fingerprint (\textit{\(RSSI_{ij}\)}) at location \textit{i}, \textit{n} is the number of access points in the database, and \textit{M} is the number of entries in the database. Once the distance between all the points in the database is calculated, the \textit{k} nearest matches are selected and the \textit{(x,y)} positions of those points averaged to obtain a final estimate of the receiver's location. This can be done as~\cite{knnformula_2}:

\begin{equation}
\label{eq:avgknn}
(x,y) = \frac{1}{k} \sum_{i=1}^{k}(x_i, y_i)
\end{equation}

\subsubsection{Naive Bayesian Estimation}
\begin{figure}[t!]
\centering
\includegraphics[width=0.6\linewidth]{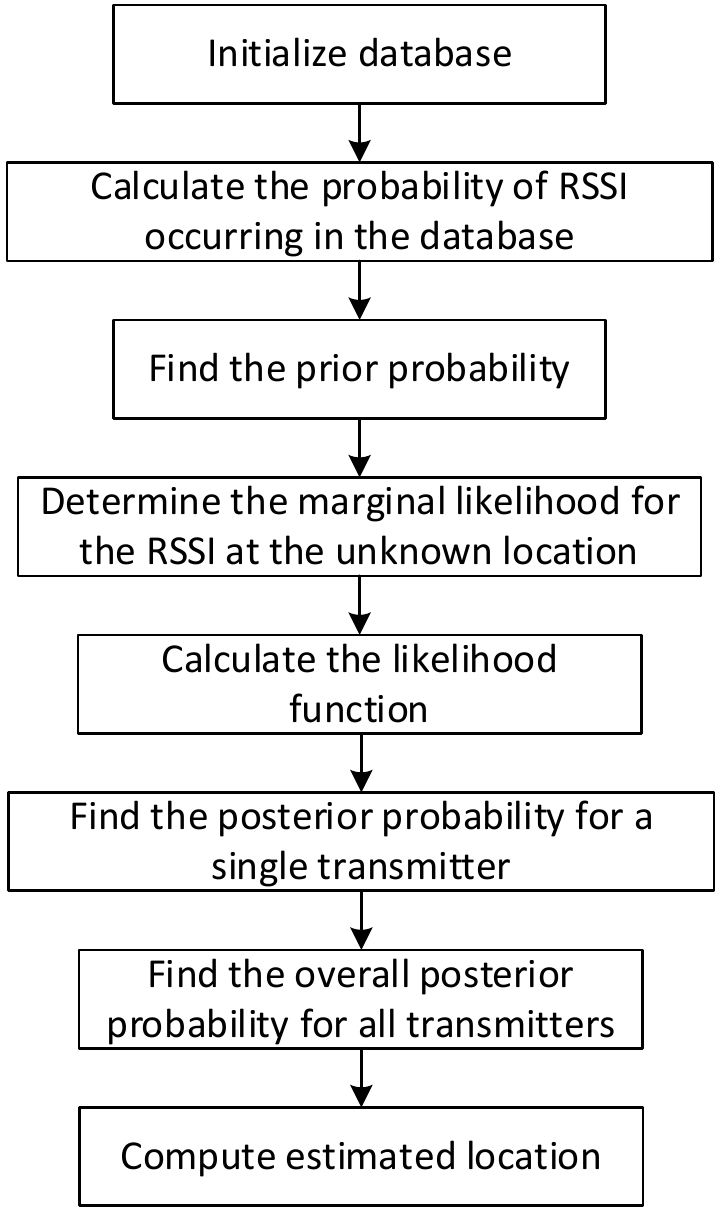}
\caption{Naive Bayes processing algorithm flowchart.}
\label{fig:bayesflowchart} 
\end{figure}

Naive Bayes is another fingerprinting algorithm that compares RSSI measurements at an unknown location with RSSI values stored in a database. Naive Bayes computes the position of the receiver by using probabilities of the stored RSSI values to determine the location with the highest likelihood of occurring. The Naive Bayes process flowchart is shown in Fig.~\ref{fig:bayesflowchart}. 

The Naive Bayes estimation is based on the Bayes Theorem: 

\begin{equation}
\label{eq:bayes}
P(A|B) = \frac{P(B|A) P(A)}{P(B)}
\end{equation}
which states that $P(A|B)$ is the posterior probability, modelling the uncertainty of event $A$ on $B$, $P(B|A)$ is the likelihood function, $P(A)$ is the prior probability, and $P(B)$ is the marginal likelihood.

In Naive Bayes, the probability of an event occurring is calculated using Eq. (\ref{eq:bayes}). The event with the highest probability is considered the most likely option to occur. By using Bayes Theorem as a base, a formula can be derived that is able to compare measured RSSI values at an unknown location to RSSI values that are stored in the database to determine the most likely position in the database of an unknown device. 

By replacing \textit{A} in Eq. (\ref{eq:bayes}) with \(y_i\) and \textit{B} with \textit{S}, then~\cite{bayes}:

\begin{equation}
\label{eq:bayes1}
P(y_i|S) = \frac{P(S|y_i) P(y_i)}{P(S)}
\end{equation}
where $y_i$ is the RSSI measurement for an access point stored at position $i$ in the database, and $S$ is the RSSI from the corresponding access point of the unknown location being mapped. In order to determine $P(S)$, the sum of the probabilities for an observed RSSI value at location $S$ is given to the database at location $y$ multiplied by the probability of the RSSI value occurring in the database. Then, $P(S)$ can be calculated as:

\begin{equation}
\label{eq:bayes2}
P(S) = \sum_{i=1}^{n}(P(S|y_i)P(y_i))
\end{equation}
where $n$ is the number of locations stored in the database. To calculate $P(S|y_i)$, all possible RSSI values are assumed to be Gaussian distributed based on the measurements that were collected during the training phase.

To determine the location that maximizes $P(y_i|S)$, for all the access points that are stored in the database, an average of probabilities can be taken. A summary of the calculations performed is:

\begin{equation}
\label{eq:bayes3}
P(y_i|S) = \frac{1}{n}\sum_{j=1}^{n} P(y_{ij}|S_j)
\end{equation}
where $n$ is the number of access points in the database and $j$ refers to a specific access point.

\subsection{Error Analysis}
To determine which localization technique produces the most accurate results, the Mean Squared Error (MSE) between the calculated position \((x_{calc}, y_{calc})\) and the real position \((x_{real}, y_{real})\) was determined,  as~\cite{knnFormula}:

\begin{equation}
\label{eq:error}
Error = \sqrt[]{(x_{calc}-x_{real})^{2} + (y_{calc}-y_{real})^{2}}
\end{equation}
 
 Once the errors for all tests performed were found, an average was taken which could then be compared to the other techniques to determine which produced results closest to the actual location. 

\subsection{Computational Complexity}
The complexity of an algorithm is an important factor to consider when selecting a technique to use for indoor localization. If the database created is large, there would be a large number of points that could be matched with producing a higher accuracy. However, this would also increase the number of computations that would need to be performed in finding the optimal match, increasing the total runtime of the algorithm. Table~\ref{tb:complexity} summarizes the computational complexity for the algorithms tested. Note that $m$ is the number of transmitting nodes and $n$ is the number of reference points in the database. 

\begin{table}[t!]
\centering
\caption{Computation complexity.}
\label{tb:complexity}
\begin{tabular}{|c|c|}
\hline
\textbf{Algorithm} & \textbf{Space complexity} \\ \hline
Trilateration & \textit{O}(1) \\ \hline
KNN & \textit{O}(mn) \\ \hline
Naive Bayes & \textit{O}(mn) \\ \hline
\end{tabular}
\end{table}

Trilateration is  one of the fastest and simplest methods available capable of functioning with complexity \textit{O}(1). Being purely sequential, trilateration simply takes RSSI values as inputs, converts them to distances and is able to calculate the estimated position of an unknown device. Unlike trilateration, both KNN and Naive Bayes require a significantly higher amount of time to calculate a position of a device which causes them to have a much larger complexity that evaluates to \textit{O}(mn). To calculate a position, KNN and Naive Bayes needed to search through a database comparing the RSSI measurements of the stored points to the unknown location performing $n$ comparisons, with $m$ possible transmitters to verify RSSI measurements with.

\section{Wireless Technologies} \label{metho}
In the design of an indoor localization system, selecting the proper wireless technology to be used is an important factor that needs to be considered. While many different types of localization systems have been proposed in the literature and developed utilizing technologies such as WiFi~\cite{wifiJournal}, Bluetooth Low Energy (BLE)~\cite{rezazadeh, bleJournal}, Zigbee~\cite{zigbeeJournal}, Radio Frequency Identification Device (RFID)~\cite{rfidExample, Carotenuto}, and Ultra Wideband (UWB)~\cite{uwbJournal}~\cite{survey1}, no standard exists and the designers to select the wireless technology following the application requirements. A comparison of the energy requirements of the different IoT technologies can be found in~\cite{mine}.

\subsection{Zigbee - IEEE 802.15.4}
In order to create a Zigbee network, Series 2 2mm Wire Antenna XBees were used as the communication devices. The XBees are small, easy to use antennas that can quickly create Zigbee networks with high-throughput and low latency. However, due to the limited processing power of the XBees, a microcontroller was necessary in order to control the flow of information. An Arduino Uno was selected due to its low power consumption and ease of integration with the XBee. 

\subsection{Bluetooth Low Energy (BLE)}

For the BLE experiments, Gimbal Series 10 Beacons were selected as transmitting devices. The Gimbal beacons were configured to use the iBeacon protocol developed by Apple which is used to characterize beacons. The iBeacon packet structure is able to define three fields, the Universally Unique Identifier (UUID), Major value, and Minor value, that are user-configurable to assist in identifying a set of beacons. For the experiments, similar UUID and Major values were set to all of the beacons, while the Minor values were altered to separate specific beacons. One benefit of using beacons is due to its one-way transmission. This allows for only the receiver to be used in tracking and not the transmitters so tampering cannot occur within the system.
To receive the signals and measure the RSSI values, a Raspberry Pi 3 Model B was used. 

\subsection{WiFi (2.4 GHz) - IEEE 802.11.N}

In order to create a Wireless Local Area Network (WLAN) using WiFi, Raspberry Pi 3 Model Bs were selected as the transmitting and receiving devices. The Pi 3s contained an onboard WiFi antenna, therefore, it allowed for a WLAN to be created. The RSSI of the WiFi signals could be easily measured by simply polling the antenna for any available signals and then focusing only on the ones of interest. 

\section{Experimental Scenarios} \label{scenarios}
To compare the performance of the techniques and the technologies under different environments, three scenarios were used, one with low, one with high and one with average interference level.

\subsection{First Scenario - Small room with low interference.} 
\begin{figure}[t!]
\centering
\subfloat[Fingerprints.]
{\includegraphics[width=.5\columnwidth]{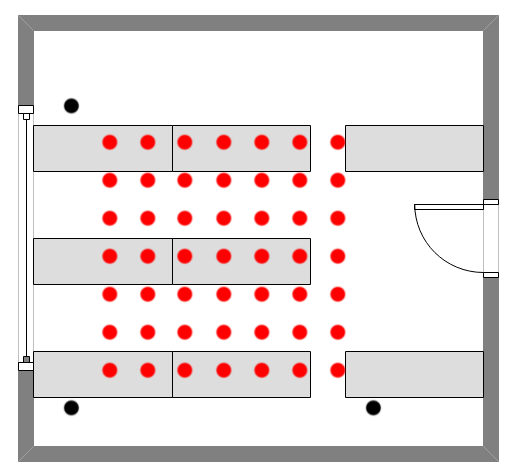}\label{fig:s1_f}}
\subfloat[Evaluation Points.]
{\includegraphics[width=.5\columnwidth]{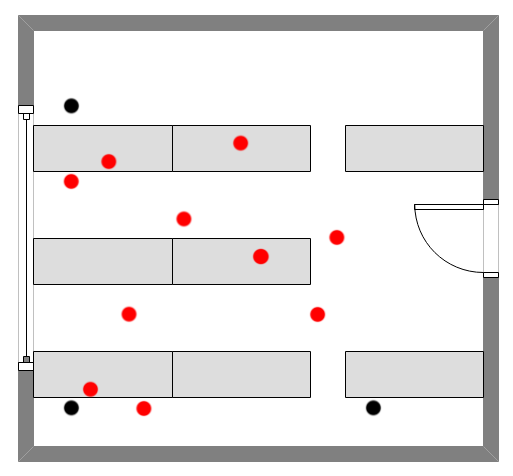}\label{fig:s1_t}}
\caption{Scenario~1 - Small room (6~$\times$~5.5~m) with low interference.}
\label{s1}
\end{figure}

The first scenario focused on an environment with low interference. A  small meeting room was selected, which contained only tables and chairs hence, there was no unnecessary interference in the area or other transmitting devices. The size of the meeting room was approximately 33~m\textsuperscript{2}, measuring 6~$\times$~5.5~m. The layout for this testbed can be seen in Fig.~\ref{s1}, with the fingerprint locations in Fig.~\ref{fig:s1_f} and the evaluation points in Fig.~\ref{fig:s1_t} . The black dots are the stationary transmitters and the red dots are the points of interest measured during the first phase of fingerprinting. 

For this environment, a dense fingerprint map was created by spacing points of interest apart by 0.5~m in a grid fashion. This was done in order to determine how the accuracy is affected when a larger number of locations with close proximities are stored in the database. In total there were 49~fingerprints obtained. The transmitters were configured in a right angle triangle where the spacing between them was 4~m apart. To test the positioning algorithms, random points in the environment were selected and the signal strength readings from each of the transmitters recorded. For this scenario, 10~points represented in red in Fig.~\ref{fig:s1_t}, were obtained to be used in calculating the accuracy of the positioning techniques.

\subsection{Second Scenario - Small room with high interference.} 
\begin{figure}[t!]
\centering
\subfloat[Fingerprints.]
{\includegraphics[angle=90,width=.5\columnwidth]{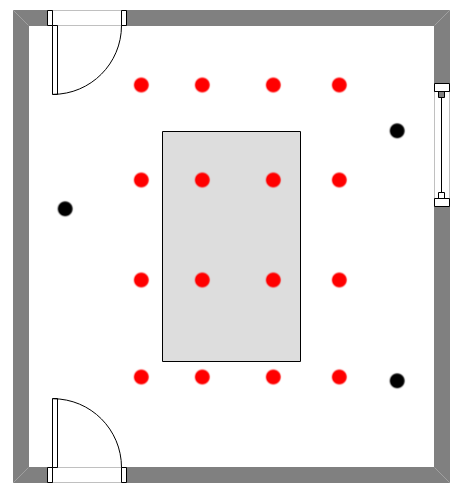}\label{fig:s2_f}}
\subfloat[Evaluation Points.]
{\includegraphics[angle=90,width=.5\columnwidth]{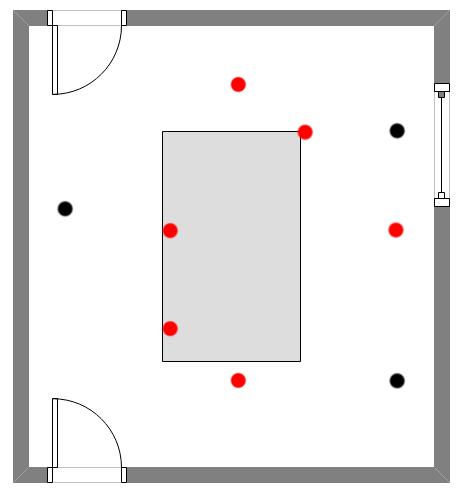}\label{fig:s2_t}}
\caption{Scenario~2 - Small room (5.8 $\times$ 5.3~m) with high interference.}
\label{s2}

\end{figure}

The second scenario focused on an environment with high interference. A small meeting room was selected and interference was created on purpose. In addition to the tables and chairs that were present in the area, four transmitting devices were distributed inside the environment. For each of the experiments, four additional transmitters configured similarly to the ones used for testing were placed in the environment. The size of the meeting room was approximately 31~m\textsuperscript{2}, measuring 5.8~$\times$~5.3~m. The layout for this testbed is shown  in Fig.~\ref{s2}, with the fingerprint locations in Fig.~\ref{fig:s2_f}. The black dots are the stationary transmitters, while the red dots are the points of interest. 

For this environment, a sparse database was created where only 16~points were obtained. The transmitters were positioned in a fashion that allowed for unique RSSI values to be measured at the different points of interest. To gather data to be used in comparing the different positioning techniques, 6~points were randomly selected to be measured. The location of the testing points can be seen in red in Fig.~\ref{fig:s2_t}.

\begin{figure}[t!]
\centering
\subfloat[Fingerprints.]
{\includegraphics[angle=90,width=.5\columnwidth]{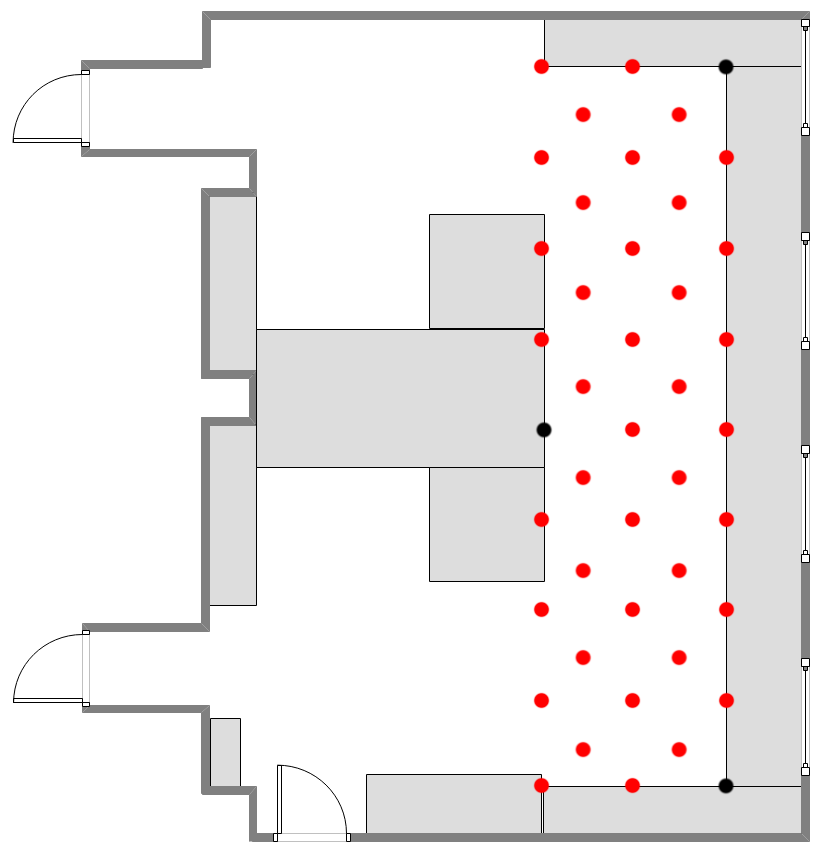}\label{fig:s3_f}}
\subfloat[Evaluation Points.]
{\includegraphics[angle=90,width=.5\columnwidth]{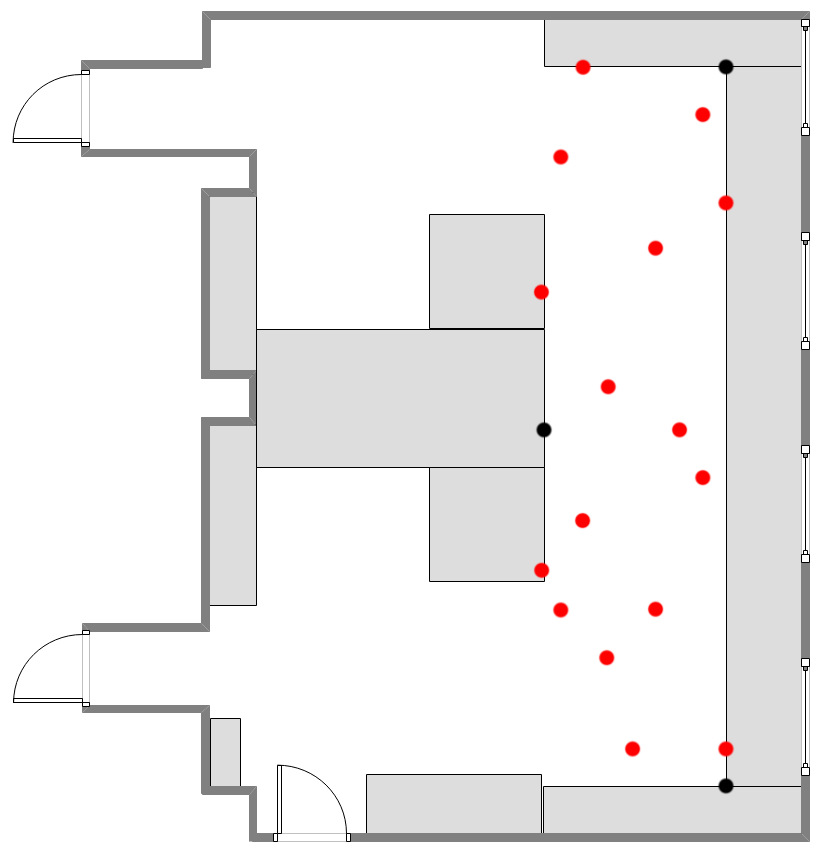}\label{fig:s3_t}}
\caption{Scenario~3 - Large laboratory (10.8 $\times$ 7.3~m) with average interference.}
\label{s3}

\end{figure}

\subsection{Third Scenario - Large room with average interference.} 
The third scenario focused on a room with an average interference due to people and wireless equipment in the area. A large computer lab was selected, which was a standard working environment and contained tables, chairs, BLE devices, and WiFi-enabled computers. Due to the large number of objects occupying this environment, it allowed for signals to experience obstructions, interference, and reflections. The size of the lab was approximately 79~m\textsuperscript{2}, measuring 10.8~$\times$~7.3~m. The layout for this testbed is shown in Fig.~\ref{s3}, with the fingerprint locations  in Fig.~\ref{fig:s3_f}. The black dots are the stationary transmitters, while the red dots are the points of interest measured during the first phase of fingerprinting. 

In this environment, only a portion of the room was utilized for fingerprinting as Line-of-Sight~(LoS) was needed to properly compare trilateration with the fingerprinting techniques. For this scenario, 40~references were gathered. The reference points were uniformly distributed between the transmitters, with an alternating pattern being used to produce a non-uniform grid. The testing data set gathered consisted of 16~points randomly selected throughout the experimental area. The location of the testing points can be seen in red in Fig.~\ref{fig:s3_t}.

\section{Experimental Procedure} \label{exp}

In order to determine which of the localization techniques produced the highest accuracy, a detailed experiment needed to be completed that would be able to utilize the strengths and weaknesses of each technique. 

\subsection{Process}

For all the experiments performed in each scenario, experiments were set up in order for LoS to be available between all the transmitters and receivers. All the nodes were placed on tables during the experiments. This was done in order for RSSI values to be recorded at a height similar to that of an individual carrying a receiving device in their pocket or a bag.

For each scenario, the initial phase of fingerprinting included reference locations consisted of 300 scans in the environment gathering approximately 100 scans from every access point. Once scanning was complete a moving average was taken of all the RSSI values from each of the access points. A similar procedure was also used in measuring the testing locations. Due to RSSI values being highly prone to interference, doing this allowed for a measurement that was better reflected by the actual signal strength than one greatly affected by interference.

\subsection{Device configuration} 
All of the transmitting devices needed to be properly configured with appropriate transmission power and transmission interval. These are important parameters as not only they affect the accuracy and response time of the system but they also affect the power consumption of the device. This becomes critical in systems where devices require limited power supplies to function as a larger power usage would cause the device to run for a shorter time.

For these experiments, the devices were set to broadcast using a transmission power of $-$~10~dBm. This was selected based on pre-configured levels for the Series 2 XBees and Gimbal Series 10 beacons. Both devices only had one value in common, hence, $-$~10~dBm was chosen. The Raspberry Pi~3 does not have the same limitation and could be dynamically configured with the chosen value. For the transmission interval, a frequency of 2~Hz was utilized corresponding to a time of 0.5~s. When selecting an appropriate transmission interval, the Gimbal beacons again had a list of pre-configurable values that could be chosen. The other devices utilized were microcontrollers, hence, not limited to a list of specific times and could be programmed with a desired value. Due to an indoor localization system requiring a short response time in the updating of locations of the targets, a quick transmit interval was necessary. Hence, a time of 0.5~s was chosen. 
 
\subsection{Path-loss model}
For trilateration, the distances between devices was found using the RSSI and path-loss model, as:
\begin{equation}
\label{eq:pathloss}
RSSI = -10 n log_{10}(d) + C
\end{equation}
where the path-loss exponent \textit{n}, and constant \textit{C}, were required for each of the wireless communication technologies in each of the testing scenarios. To determine the corresponding values, a single transmitter and a receiver were set up where the RSSI could be measured for a range of distances.

\begin{figure*}[t!]          
\centering
\subfloat[Scenario 1.]
{\includegraphics[width=.65\columnwidth]{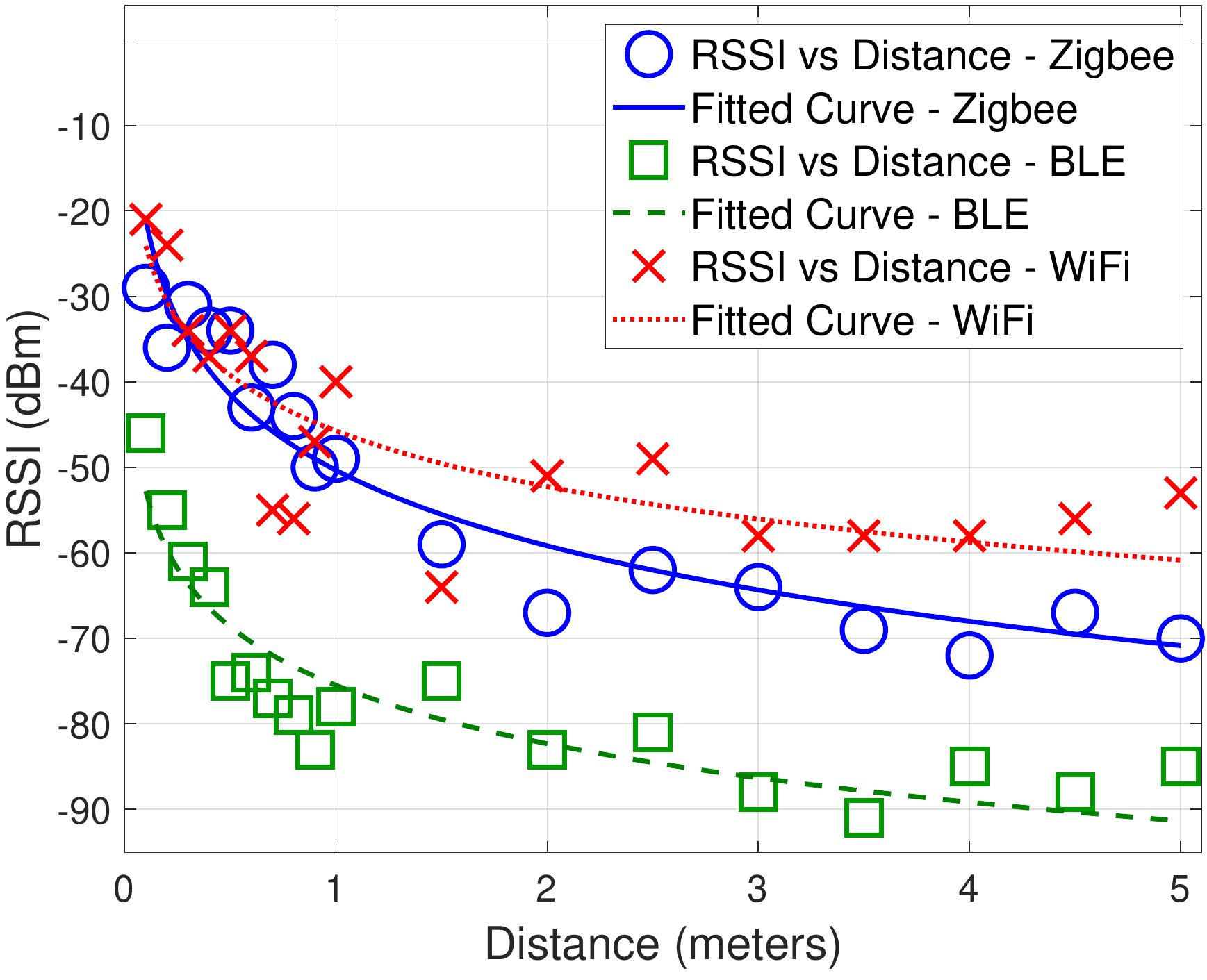}\label{sc1path}}
\hfill
\subfloat[Scenario 2.]
{\includegraphics[width=.65\columnwidth]{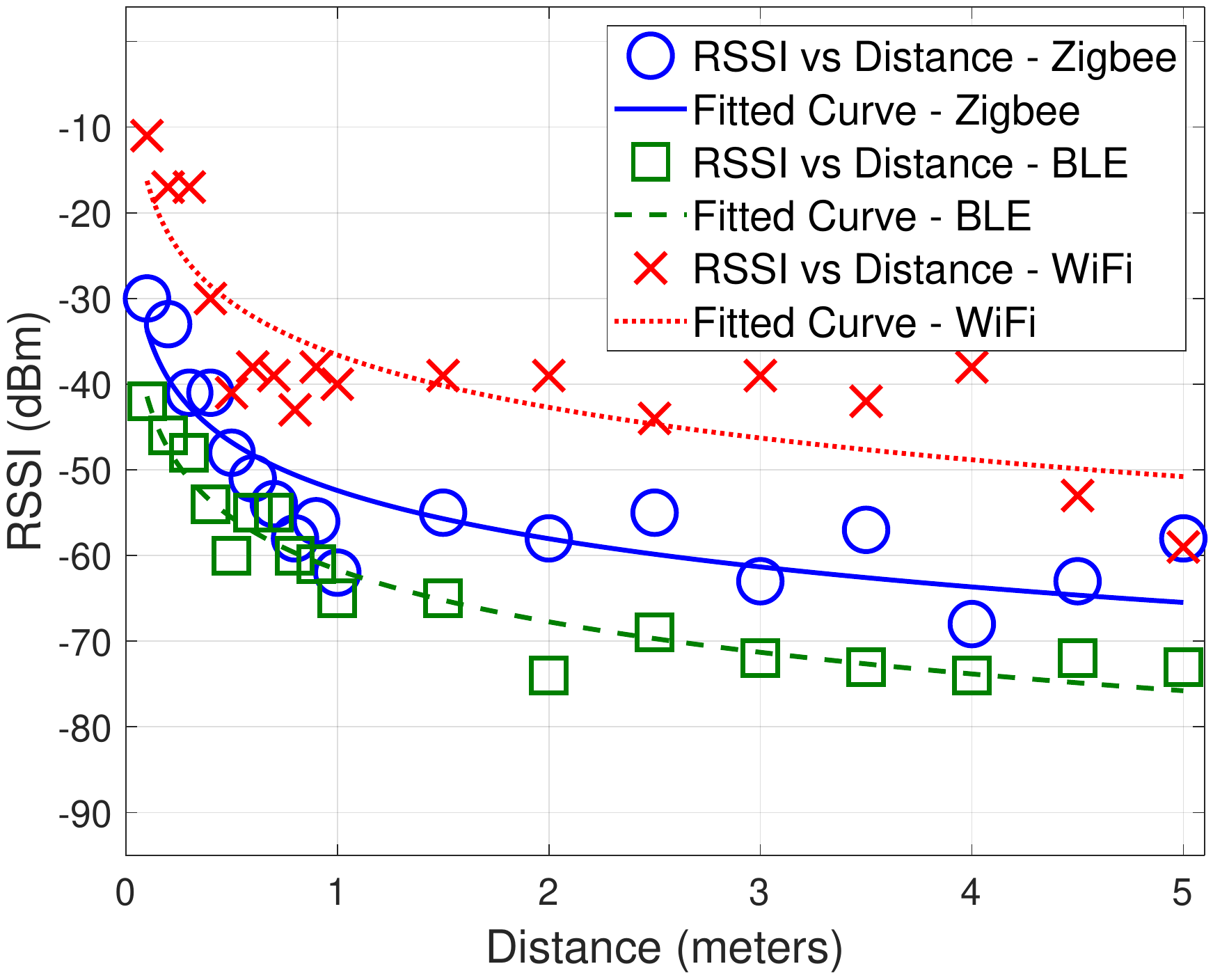}\label{sc2path}}
\hfill
\subfloat[Scenario 3.]
{\includegraphics[width=.65\columnwidth]{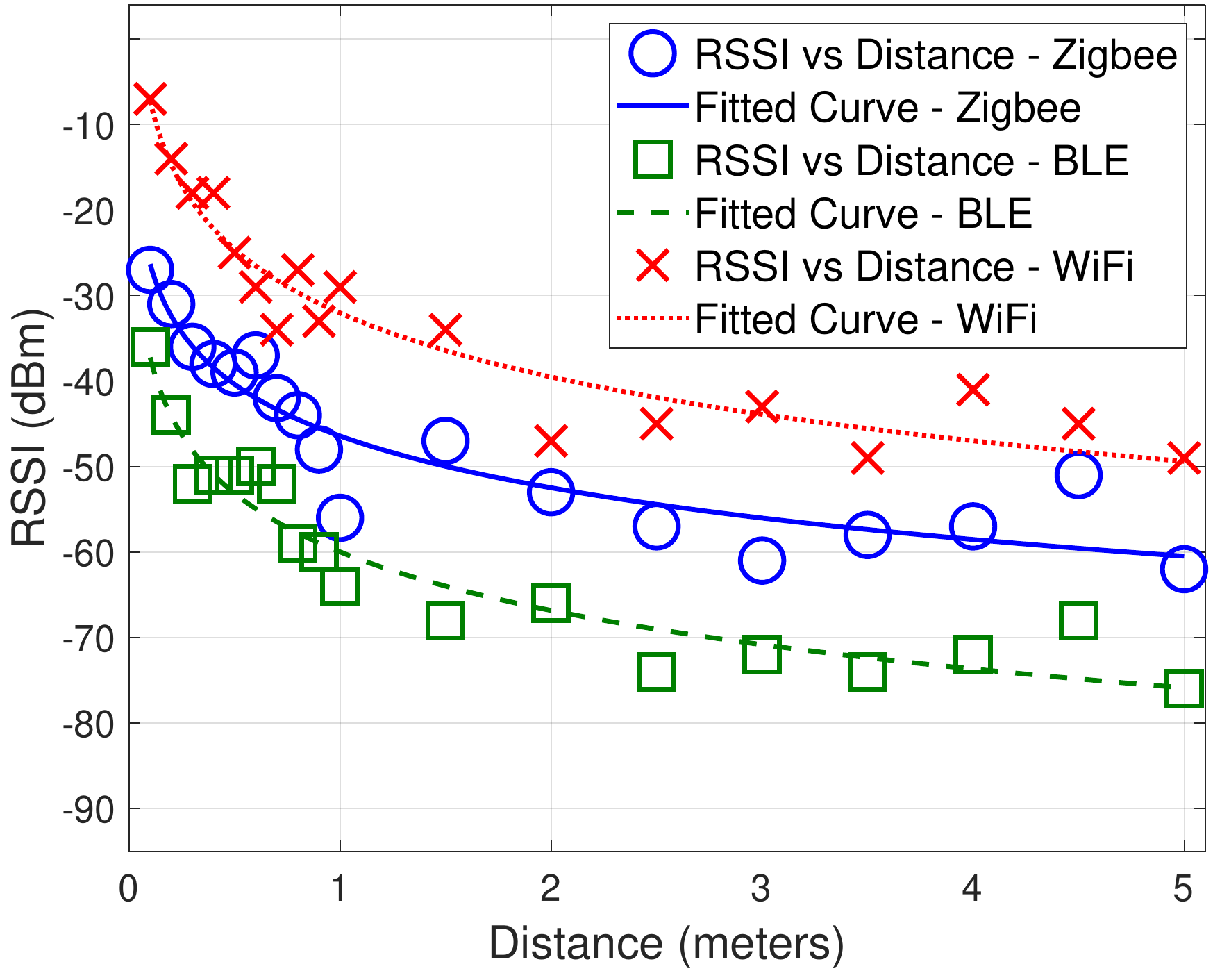}\label{sc3path}}
\caption{Curve fitting models for the three different scenarios.}
\label{fig:models3}
\end{figure*}

For our purposes, the RSSI was measured eighteen times at varying distances, every 0.1~m from 0 to 1~m, and every 0.5~m from 1 to 5~m. Due to a signal losing most of its strength once it is transmitted, a larger number of points were taken at shorter distances than further ones. The recorded distances and the measured RSSI values were then plotted.  After using a curve fitting function, models were created to determine the required parameters for each of the testing environments. The resulting models from performing the curve-fitting can be seen in Fig.~\ref{sc1path}, Fig.~\ref{sc2path}, and Fig.~\ref{sc3path} for Scenarios 1, 2, and 3, respectively. The parameters determined from the models created for each of the wireless communication technologies for each of the experimental scenarios can be seen in Table~\ref{tb:pathloss}.

In order for RSSI measurements to be consistent, the transmitters and receiver were positioned in the same orientation throughout the tests. Since the direction the device faces can have a significant impact on the RSSI readings~\cite{orientation}, the orientation was kept the same while generating the fingerprint database and collecting test points. This would allow for a smaller error to occur as all the signals would have the same distance to travel in order to reach the device's antenna. In an attempt to minimize the error, devices were placed in the center of the data collection point to ensure that the correct distance was maintained at all times. 
   
\begin{table}[t!]
\centering
\caption{Parameters obtained to convert RSSI into distance using the path-loss model.}
\label{tb:pathloss}
\begin{tabular}{|c|c|c|c|}
\hline
 \textit{parameter}& \textbf{Zigbee} & \textbf{BLE} & \textbf{WiFi} \\  \hline \hline
\multicolumn{4}{|c|}{\textbf{Scenario~1}} \\ \hline
\textit{n} & 2.935 & 2.271 & 2.162 \\ \hline
\textit{C} & -50.33 & -75.48 & -45.73 \\ \hline
\textit{R\textsuperscript{2}} & 0.9051 & 0.85 & 0.7177 \\ \hline \hline
\multicolumn{4}{|c|}{\textbf{Scenario~2}} \\ \hline
\textit{n} & 1.912 & 1.999 & 2.018 \\ \hline
\textit{C} & -52.73 & -62.27 & -37.37 \\ \hline
\textit{R\textsuperscript{2}} & 0.7689 & 0.9274 & 0.7091 \\ \hline \hline
\multicolumn{4}{|c|}{\textbf{Scenario~3}} \\ \hline
\textit{n} & 2.085 & 2.442 & 2.563 \\ \hline
\textit{C} & -48.52 & -62.5 & -33.75 \\ \hline
\textit{R\textsuperscript{2}} & 0.9006 & 0.9317 & 0.9294 \\ \hline
\end{tabular}
\end{table}

\subsection{Techniques configuration}
Simple trilateration has poor performance when it is applied to raw data. To improve the performance of trilateration, Kalman filter was applied to the raw data before they are used with this technique.

When using KNN algorithm for fingerprinting the proper selection of $k$ value is crucial. In~\cite{landmarc}, fingerprinting experiments were performed using a KNN algorithm, where tests were done in order to find the $k$ value that produces the best results. Through the tests performed, it was determined that $k = 4$ produced the best results. After repeated experiments performed at later dates, it was found that $k = 4$ still produced values with the highest accuracy. We also experiment with different values through all the data in the dataset and we found that $k = 4$ would produce the best results and be optimal for our experiments as well.

\section{Results and Discussion} \label{res}
In this section, we present the results for the experiments performed along with a detailed discussion on what the results signify. The experimental data are  available online at~\cite{RSSIdata}.

\subsection{Experimental Results}
\begin{figure*}[t!]          
\centering
\subfloat[Scenario 1.]
{\includegraphics[width=.7\columnwidth]{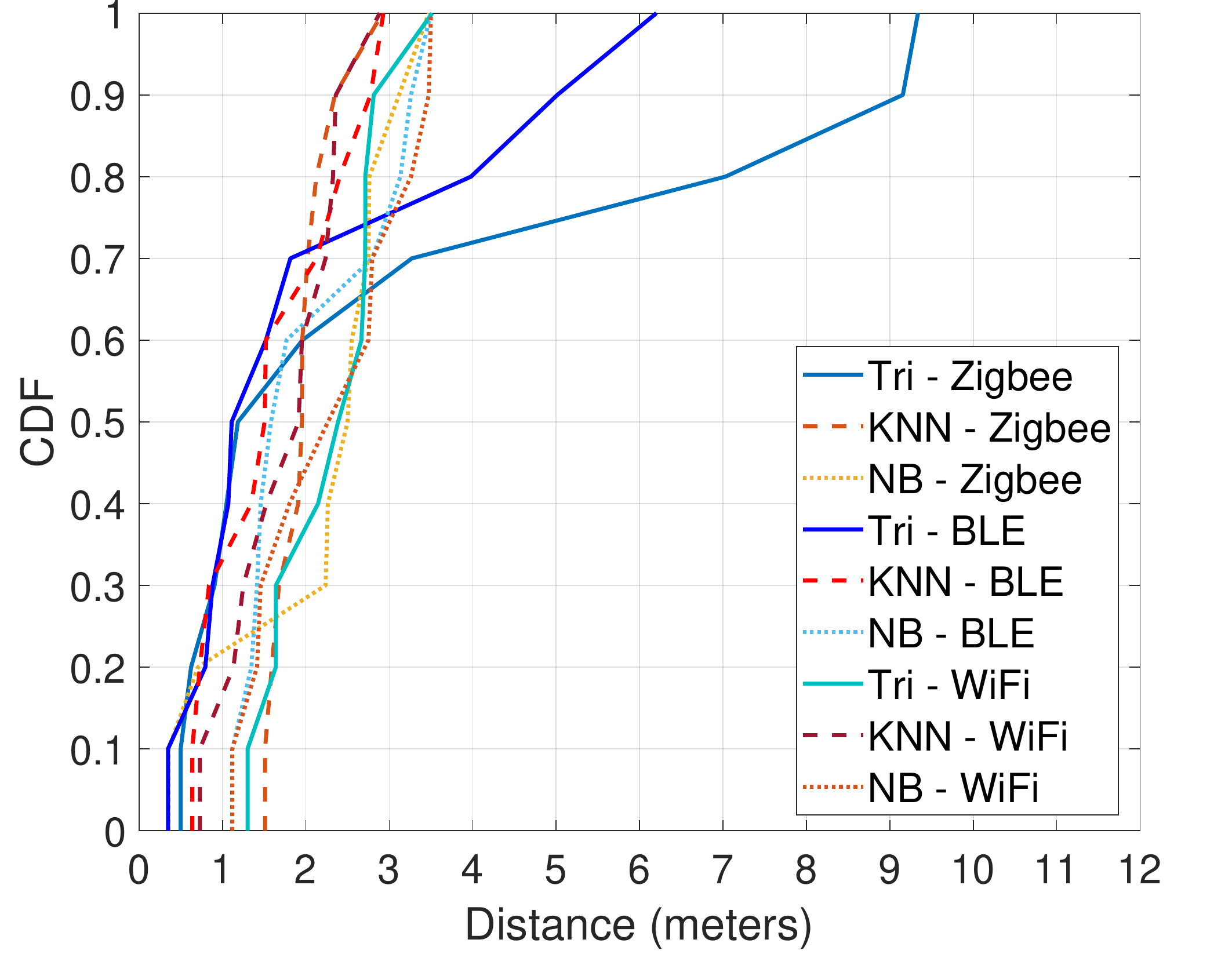}\label{cdf1}}
\subfloat[Scenario 2.]
{\includegraphics[width=.7\columnwidth]{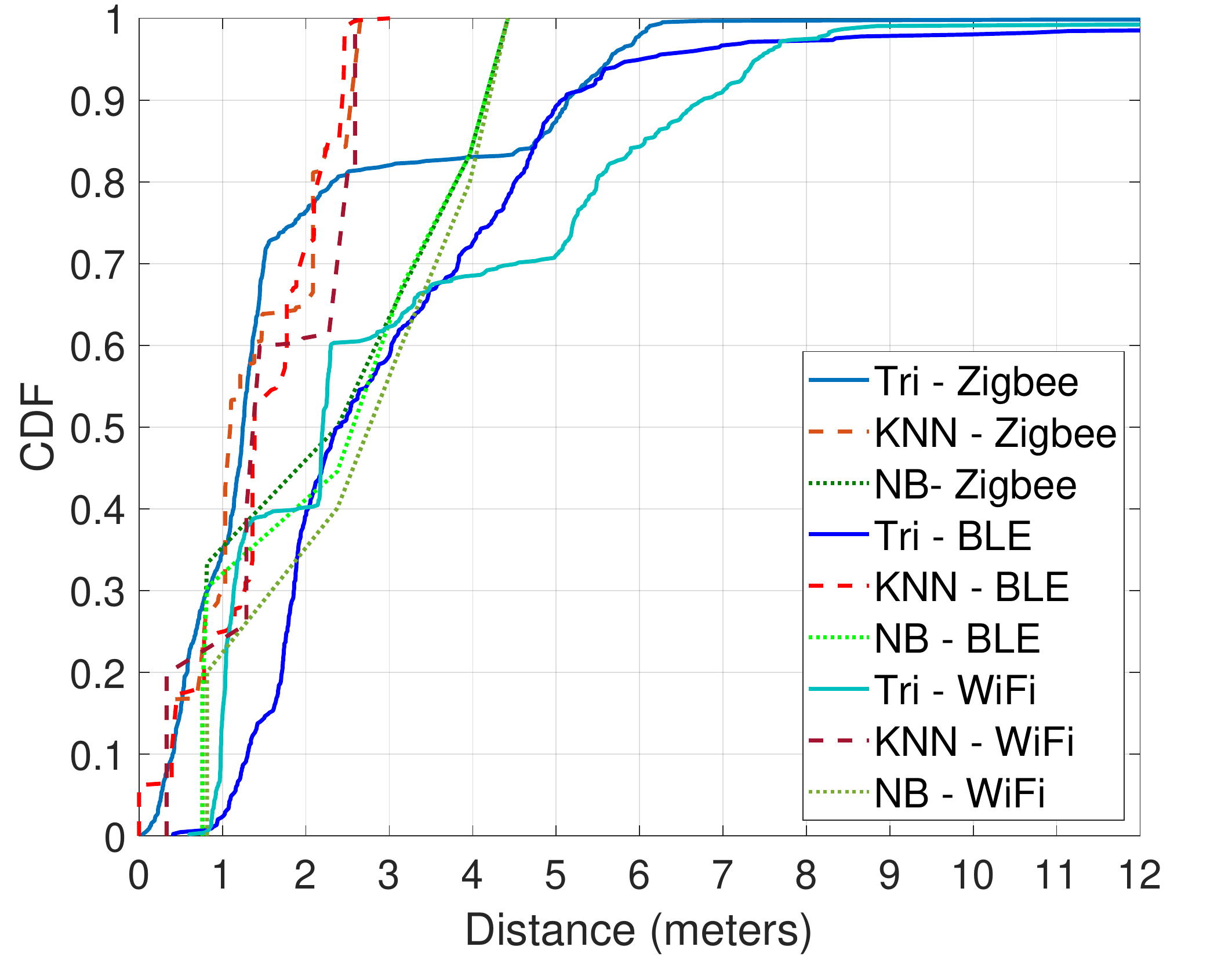}\label{cdf2}}
\subfloat[Scenario 3.]
{\includegraphics[width=.7\columnwidth]{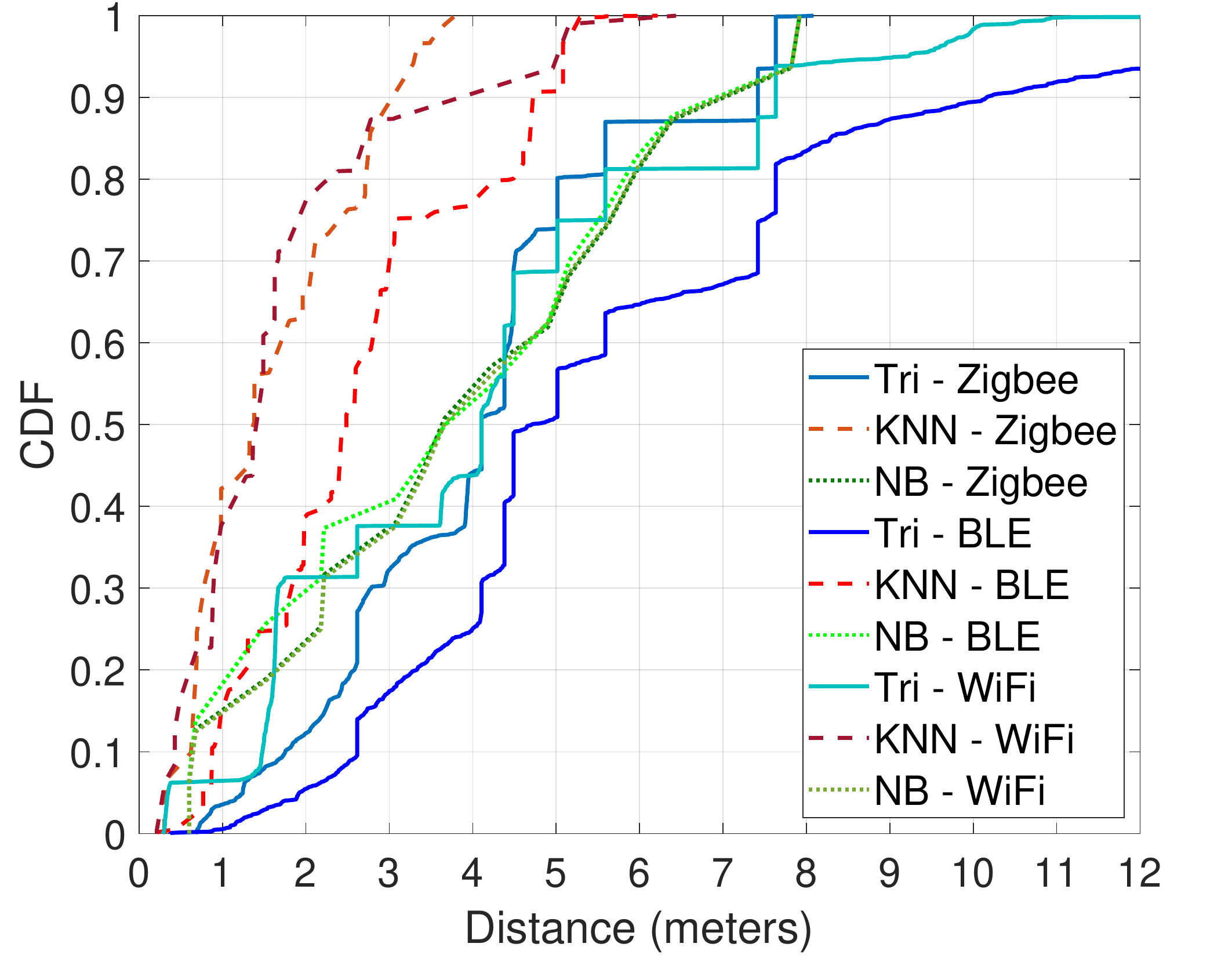}\label{cdf3}}
\caption{CDF for the three scenarios.}
\label{cdf_all}
\end{figure*}

The overall cumulative error for the techniques and the technologies for each scenario is shown in Fig.~\ref{cdf_all}. A summary of the experimental results for the tests performed can be seen in Table~\ref{tb:results} for the MSE calculations along with the variance. 

The CDF results clearly show that KNN performs better than the other two approaches for all three scenarios. For Scenario 1, shown in Fig.~\ref{cdf1}, KNN has an error less than 2.5~m 95\% of the time, followed by Naive Bayes with an error less than 3.5~m 95\%. For Scenario 2, shown in Fig.~\ref{cdf2}, KNN has again the best performance with an error less than 2.8~m 95\% of the time. Finally, for Scenario 3, KNN has an error less than 5.1~m 95\% of the time. In terms of the technologies, BLE has the best performance in Scenario~1 and Scenario~2, followed by WiFi and Zigbee, respectively. For Scenario~3, the best performance comes from WiFi, followed by BLE.
 
The MSE values and variances are shown in Table~\ref{tb:results}. According to the experimental results, KNN, with $k = 4$, produced calculations with the greatest overall accuracy and precision. The estimates computed using KNN deviated off of the actual receiver by 1.602~m with a computed variance of 0.6662~m over all the rooms and the examined technologies. KNN was also the technique that produced the lowest errors in each of the experimental scenarios utilized. In Scenario~1, the average error was 1.8376~m with a variance of 0.4375~m. In Scenario~2, the average error was 1.3581~m with a variance of 0.6759~m. In Scenario~3, the average error was 1.6104~m with a variance of 0.885~m.

\begin{table*}[t!]
\centering
\caption{Summary of error and variance for localization techniques and wireless technologies (meters).} 
\label{tb:results}
\begin{tabular}{|c|c|c|c|c|c|c|c|c|}
\hline 
\textbf{\begin{tabular}[c]{@{}c@{}}\textit{Positioning} \\ \textit{Technique}\end{tabular}} & \multicolumn{2}{c|}{\textbf{Trilateration}} & \multicolumn{2}{c|}{\textbf{KNN}} & \multicolumn{2}{c|}{\textbf{Naive Bayes}}  & \multicolumn{2}{c|}{\textit{Average}}\\ \hline \hline
& MSE & $\sigma^2$& MSE & $\sigma^2$& MSE & $\sigma^2$& MSE & $\sigma^2$ \\\hline
\hline
\multicolumn{9}{|c|}{\textbf{Scenario~1}} \\ \hline
Zigbee 		& 3.5016 			& 7.1631		& 2.0007 	& 0.1672	& 2.3068  		& 1.0668		&  \textit{2.603} &2.799\\ 
BLE 			&  \textbf{2.2734} 			&4.1409 		&\textbf{1.6814} 	&0.7059	 		&2.3834 		&0.8061		& \textit{2.1127} &1.8843\\ 
WiFi 			&2.3529 	&0.4536 		&1.8307 	&0.4394 			&\textbf{2.1372} 		&0.8483		& \textbf{\textit{2.1069}}& 0.5798\\ \hline \hline
\textit{Average} & \textit{2.7093} 	&3.9192		&\textbf{\textit{1.8376}} 	& 0.4375		&\textit{2.2758} &0.9071 & 2.2742 &  1.7546 \\ 
\hline
\hline
\multicolumn{9}{|c|}{\textbf{Scenario~2}} \\ \hline
Zigbee & \textbf{2.0851} &5.4278& 1.3803 &0.7377& 2.5814 &2.4208&  \textbf{\textit{2.0156}}&2.8621\\ 
BLE & 2.9406 &2.1079& \textbf{1.2794} & 0.7089 & 2.1905 &2.04&  \textit{2.1368}& 1.6189\\ 
WiFi & 3.4214 &6.0842& 1.4147 &0.5812& \textbf{2.0717} &2.0715&  \textit{2.3026}&2.9123\\ \hline \hline
\textit{Average} & \textit{2.8157} &4.54  &\textbf{\textit{1.3581}} & 0.6759& \textit{2.2812} &2.1774& 2.1517 & 2.4644 \\ \hline \hline

\multicolumn{9}{|c|}{\textbf{Scenario~3}} \\ \hline
Zigbee & 4.2323&5.4876 & 1.7984 &1.3857& 3.8673 &5.2735&  \textit{3.2993}& 4.0489\\ 
BLE & \textbf{3.7358}&3.9799 & 1.6472 &0.6773& \textbf{2.739} &4.091&  \textbf{\textit{2.7073}}&2.9161\\ 
WiFi & 4.7907&1.5336 & \textbf{1.3856} & 0.5921 & 2.8274 &3.5895&  \textit{3.0012}& 1.9051\\  \hline \hline
\textit{Average} & \textit{4.2529}&3.667 & \textbf{\textit{1.6104}} & 0.885 & \textit{3.1446} &4.318&  3.0026 & 2.9565\\ \hline \hline
\textbf{Overall} & \textbf{3.2593} &4.0421& \textbf{1.602} &0.6662& \textbf{2.5672} &2.4675& 2.4762 & 2.3919\\ \hline
\end{tabular}
\end{table*}
 
No single wireless technology utilized produced results that greatly affected the system calculations. In Scenario~1, BLE produced results with the lowest error of 1.6814~m, however, calculated results also had the largest variance of 0.7059~m. In Scenario~2, BLE was again the most accurate technology producing a result of 1.2794~m. Zigbee however, was second in precision in Scenario~2 with an error of 1.3803~m. In Scenario~3, WiFi proved to be the most accurate with an error of 1.3856~m. WiFi was also the most precise in Scenario~3 with a variance of 0.5921~m.

The algorithm with the next highest accuracy and precision was Naive Bayes. Overall, the estimates it produces deviated off of the actual receiver position by 2.5672~m with a variance of 2.4675~m. Based on the tests performed, Naive Bayes was not greatly affected by the size of the database. In Scenario~1 where 49 fingerprints were taken, the largest amount out of any environment, an average error of 2.2758~m was determined with a variance of 0.9071~m. In Scenario~2, where 16 fingerprints were taken, the lowest amount, a slightly larger error occurred of 2.2812~m with a much larger variance of 2.1774~m. In Scenario~3, it produced the largest errors overall for this system resulting in values that deviated by 3.1446~m with a variance of 4.318~m.

Finally, the algorithm with the worst overall accuracy and precision was trilateration. It proved to be the most unreliable and unpredictable technique for an indoor localization system. In Scenario~1, an average error for all the technologies of 2.7093~m with a variance of 3.9192~m was calculated. In Scenario~2, the average error increased to 2.8157~m with the variance increasing to 4.54~m. In Scenario~3,  the average error was increased to 4.2529~m with a reduction in the variance to 3.667~m.  Overall, trilateration deviated off of the actual receiver position by 3.2593~m with a variance of 4.0421~m. Trilateration produced the worst results in all the scenarios. 

For the selection of the proper technology, it is important to consider, not only the average MSE but also the variance.
Some useful insights come from the variance. For Scenario~1, although WiFi and BLE have similar average MSE, WiFi has a small variance of 0.5798~m versus the 1.8843~m of BLE. For Scenario~2, BLE and Zigbee have similar average MSE, however, BLE has a small variance of 1.6189~m in comparison with the 2.8621~m of Zigbee. For Scenario~3, BLE has the lowest MSE on average however, WiFi which has slightly higher MSE, it has the lowest variance.

\subsection{Discussion}
According to experimental results, when comparing different types of indoor localization algorithms in the selected environments, KNN proved to be the most accurate approach, with Naive Bayes in second, and trilateration in last. The results determined are as expected due to KNN and Naive Bayes both being fingerprinting techniques which were able to compute the location of the device based on the RSSI values stored in a database. The estimated location must have existed inside of the database which greatly limited the error that could occur using the algorithms. As a result, the errors produced were low, with a low variance indicating  high precision in the results. Trilateration did not suffer the same limitations and could estimate any possible location by knowing the location of transmitters and the path-loss  for the environment. 

Not only was KNN more accurate than Naive Bayes, but it was also much simpler to utilize when calculating a location. KNN simply required going through the data and computing the Euclidean distance between the test location and every point in the database. Once done, the \textit{k} points with the lowest distances could be averaged to find the estimated location. The main advantage of using fingerprinting with KNN is that if additional access points were to be added, the calculations could easily be modified to integrate the new data after a new database was created. When using Naive Bayes and a change occurs in the database, the probabilities for each of the stored RSSI values would need to be re-computed which would require a significantly longer amount of time to run. 

However, while being the most accurate and precise, KNN has a number of disadvantages that need to be considered. KNN requires the use of a database before the algorithm can be utilized. The runtime of the algorithm is also heavily dependent on the number of testing points taken and the number of access points utilized. If the size of the database were to greatly increase, the number of computations would also, therefore creating a longer runtime. When creating the database, the space that is chosen between the reference points is an important factor that needs to be considered. When fingerprinting with a smaller distance in between points, a greater accuracy could be achieved at the cost of system runtime. However, if points are taken too close together, the measured RSSI from the access points might not  reflect correctly the change in points, hence, multiple entries would exist in the database with similar RSSI readings. This in turn could hinder the overall performance. 

Trilateration was the worst technique tested, not only was it inaccurate, but it was also very inconsistent producing large variances in the values. RSSI measurements obtained beyond the scope of the model caused a large distance error to occur using the path-loss model. Based on complexity, trilateration was the fastest algorithm to execute. The process required to compute the location was very basic, with a simple conversion of RSSI to distance, then substituting the values into a set of formulas to get the estimated location. Some of the benefits of using trilateration for indoor localization come from the simpleness of implementing, since no database needs to be created beforehand and estimated locations are not limited to reference points in a database. 

For each  localization algorithm, the one which resulted in the highest accuracy also had the highest precision. The same could be said for the opposite where the localization algorithm with the lowest accuracy achieved the poorest precision. However, this does not  depend on the wireless technology. For KNN in Scenario~1, Zigbee was found to have the worst accuracy, however, it  has the highest precision.

When the results across the different testing scenarios compared, some interesting observations can be made. For trilateration, it was found that the errors produced in Scenario~1 and Scenario~2, which were the small meeting rooms, had better results than those in Scenario~3, the research lab. However, while the error was largest for Scenario~3, the variance produced was similar to Scenario~2 and lower than Scenario~1. The results in Scenario~1 can be attributed due to multipath effects that occur due to the small space. Since Scenario~3 was much larger, reflections off of walls did not affect the system the same way as in Scenario~1. Scenario~2 also saw additional transmitting devices placed in the area that would cause additional interference. While the errors between Scenario~1 and Scenario~2 do appear similar, the variances appeared to be different, which can conclude that the interference in the environment played a roll in the poor performance of the system.
 
When the results are compared for KNN between the different testing scenarios, a similar error between all the different system configurations was determined. Not only were the errors more consistent, but they were also much more accurate compared to the trilateration results. This can be attributed to the fingerprinting that is performed in the creation of the database. Since the points of interest in the environment are scanned and recorded, the amount of noise and interference between the different environments had no major effect on the outcome of the results. Interestingly, when KNN is used, Scenario~1 contained 49~fingerprints in the database and produced the largest error. Scenario~3 contained 40~fingerprints and produced the second largest error. Finally, Scenario~2 only contained 16~fingerprints and produced the lowest error. Therefore, the accuracy of the system is also related  to the number and the position of fingerprints in the database. According to our results, the scenario with the smallest database but a proper placement and distance between the point of interest produces the best results. On the other hand, while the accuracy was not optimal in Scenario~1, which has the most point of interest, the variance produced was the lowest out of the three environments.

\section{Conclusion} \label{con}
In this paper, we compared two memoryless  techniques KNN, and Naive Bayes, as well as trilateration to be used at an indoor localization system. Experiments were conducted in three rooms with different levels of interference and the techniques were compared in terms of accuracy, precision, and complexity. In order to verify results experiments were performed  with three wireless technologies: Zigbee, BLE, and WiFi. Results demonstrated that KNN with $k = 4$ was the most accurate and precise localization technique overall,  followed by Naive Bayes. Both KNN and Naive Bayes were found to have high run times requiring some time to perform calculations using a database, executing with complexity \textit{O}(mn). Trilateration being the worst algorithm overall, had the best complexity of \textit{O}(1), requiring very little running time to calculate a location. 

The experimental results can be used as an indicator for the selection of a proper indoor localization technique in smart buildings. The RSSI dataset created during these experiments has been made open-source~\cite{RSSIdata}. 

\bibliographystyle{IEEEtran}
\bibliography{ref}

\end{document}